\documentclass[11pt,twoside,a4paper,cr,final]{cms-tdr}
\usepackage{graphicx}

\def\svnVersion{145374:145378MP}

\begin{document}

\hyphenation{had-ron-i-za-tion}
\hyphenation{cal-or-i-me-ter}
\hyphenation{de-vices}

\RCS$Revision: 138024 $
\RCS$HeadURL: svn+ssh://svn.cern.ch/reps/tdr2/papers/XXX-08-000/trunk/XXX-08-000.tex $
\RCS$Id: XXX-08-000.tex 138024 2012-07-19 04:04:00Z alverson $
\newlength\cmsFigWidth
\ifthenelse{\boolean{cms@external}}{\setlength\cmsFigWidth{0.85\columnwidth}}{\setlength\cmsFigWidth{0.4\textwidth}}
\ifthenelse{\boolean{cms@external}}{\providecommand{\cmsLeft}{top}}{\providecommand{\cmsLeft}{left}}
\ifthenelse{\boolean{cms@external}}{\providecommand{\cmsRight}{bottom}}{\providecommand{\cmsRight}{right}}
\cmsNoteHeader{2012/178} 
\title{Measurement of isolated photon production in pp and PbPb collisions at $\sqrt{s_{_{\rm NN}}}=2.76$ TeV with CMS}

\author{Yen-Jie Lee on behalf of the CMS Collaboration}

\address{CERN, European Organization for Nuclear Research, Geneva, Switzerland}

\date{\today}

\abstract{
Isolated photon production is measured in pp and PbPb collisions at nucleon-nucleon centre-of-mass energies of 2.76~TeV in the pseudorapidity range $|\eta|<$~1.44 and transverse energies $E_{\rm T}$ between 20 and 80 GeV with the CMS detector at the LHC. The measured $E_{\rm T}$ spectra are found to be in good agreement with next-to-leading-order perturbative QCD predictions. The ratio of PbPb to pp isolated photon $E_{\rm T}$-differential yields, scaled by the number of incoherent nucleon-nucleon collisions, is consistent with unity for all PbPb reaction centralities.
}

\hypersetup{%
pdfauthor={CMS Collaboration},%
pdftitle={Measurement of isolated photon production in pp and PbPb collisions at $\sqrt{s_{_{\rm NN}}}=2.76$ TeV with CMS},%
pdfsubject={CMS},%
pdfkeywords={CMS, physics, software, computing}}

\maketitle 

Prompt photons with high transverse energy ($E_{\rm T}$) in hadronic collisions are produced directly from the hard scattering of two partons. Measured photon production cross sections provide a direct test of perturbative quantum chromodynamics (pQCD)~\cite{JETPHOX}, and constrain the proton~\cite{Ichou:2010wc} and nuclear~\cite{Arleo:2011gc} parton distribution functions (PDFs). In the case of nuclear collisions, jets are significantly suppressed~\cite{Chatrchyan:2011sx,Atlas:2010bu,chatrchyan:2012nia} but direct photons, W and Z bosons~\cite{Chatrchyan:2011ua,ATLAS:2010px,Chatrchyan:2012nt} are unaffected by the strongly interacting medium produced in the reaction. The comparison of photon production cross sections in pp and PbPb collisions allows one to estimate possible modifications of the nuclear parton densities with respect to nucleon PDFs. However, the measurement of inclusive photon production is complicated by the presence of a large background from the electromagnetic decays of neutral mesons. Backgrounds from these decays and the fragmentation photons are typically suppressed by imposing isolation requirements on the reconstructed photon candidates. The inclusive isolated photon measurements presented here are based on samples collected in pp and PbPb collisions at 2.76TeV with the Compact Muon Solenoid (CMS) detector~\cite{JINST}. The total data sample corresponds to an integrated luminosity of 231 nb$^{-1}$ and $6.8\,\mu\mathrm{b}^{-1}$
for pp and PbPb, respectively.

Particles produced in the collisions are reconstructed in the CMS detector which is detailed in ~\cite{JINST}.
The central tracking system comprises silicon pixel
and strip detectors that allow for the reconstruction of charged particles in the pseudorapidity
range $|\eta| < 2.5$, where $\eta = -\ln [\tan(\theta/2)]$
and $\theta$ is the polar angle relative to the counterclockwise
beam direction. Electromagnetic and hadron calorimeters are located outside the silicon tracking system and provide
coverage for $|\eta| < 3$. 
The calorimeters and tracking systems are located within the 3.8 T magnetic field of the super-conducting solenoid. 
CMS includes a hadron forward steel/quartz-fibre Cherenkov calorimeter, which covers the forward rapidities $3<|\eta|<5.2$ and is used to determine the degree of overlap (``centrality'') of the two colliding Pb nuclei. A set of scintillator tiles, the beam scintillator counters (BSC), is mounted on the inner side of the hadron forward calorimeter for triggering and beam-halo rejection for both pp and PbPb collisions.

To study electron rejection and the photon selection efficiency in PbPb collisions,
$\gamma$+jet, dijet, and $W\rightarrow e\nu$ events are simulated using the {\sc PYTHIA} generator (version 6.422, tune D6T)~\cite{Sjostrand:2006za}, modified
to take into account the isospin of the colliding Pb ion~\cite{Lokhtin:2005px}. These {\sc PYTHIA} events, propagated through the CMS detector using the {\sc GEANT4} package~\cite{geant4} to simulate the detector response, are embedded in minimum-bias (MB) PbPb events in order to study the effect of the underlying event (UE). The embedding is done by mixing the {\sc GEANT4} simulated digital information with the recorded MB PbPb data. These mixed samples (denoted ``{\sc{pythia+data}}") are used for signal shape studies, and for energy and efficiency corrections. At the generator level, an isolation cone of radius $\Delta R$ = $\sqrt{(\Delta\eta)^2+(\Delta\phi)^2}<0.4$ around the photon candidate direction is defined. A photon is considered to be isolated if the sum of the $E_{\rm T}$ of all the other final state particles produced from the same hard scattering inside the isolation cone is smaller than 5 GeV.

The photon reconstruction algorithm and isolation requirements in pp and PbPb collisions are detailed in Ref.~\cite{PhysRevD.84.052011} and~\cite{CMSpbpbPhoton}. The selected photon candidates are required to be within $|\eta^\gamma|<1.44$, not to match with any electron candidates, and to have $E^\gamma_\mathrm{T} > 20$ GeV. A first rejection of neutral mesons mimicking a high-$E_{T}$ photon candidate is done using the $H/E$ ratio defined as the ratio of hadronic to electromagnetic energy inside a cone of $\Delta R = 0.15$ with respect to the direction of the photon candidate~\cite{CMSppPhoton}. Photon candidates with $H/E < 0.2$ are selected for this analysis. To measure the isolation of a given photon candidate in a PbPb event, the signals detected by calorimeters and tracking systems in a cone of radius $\Delta R=0.4$ with respect to the centroid of the cluster is used. The average value of the energy deposited per unit area in the $\eta-\phi$ phase space is estimated within a rectangular region $2\Delta R$-wide and centered on $\eta^\gamma$ in the $\eta$-direction and 2$\pi$ wide in the $\phi$-direction, excluding the isolation cone. The UE-subtracted transverse energy in the isolation cone calculated from calorimeter and tracks with $p_T>2$ GeV/$c$ is required to be smaller than 5 GeV. 

\begin{figure*}[t]
  \begin{center}
    \includegraphics[width=0.95\textwidth]{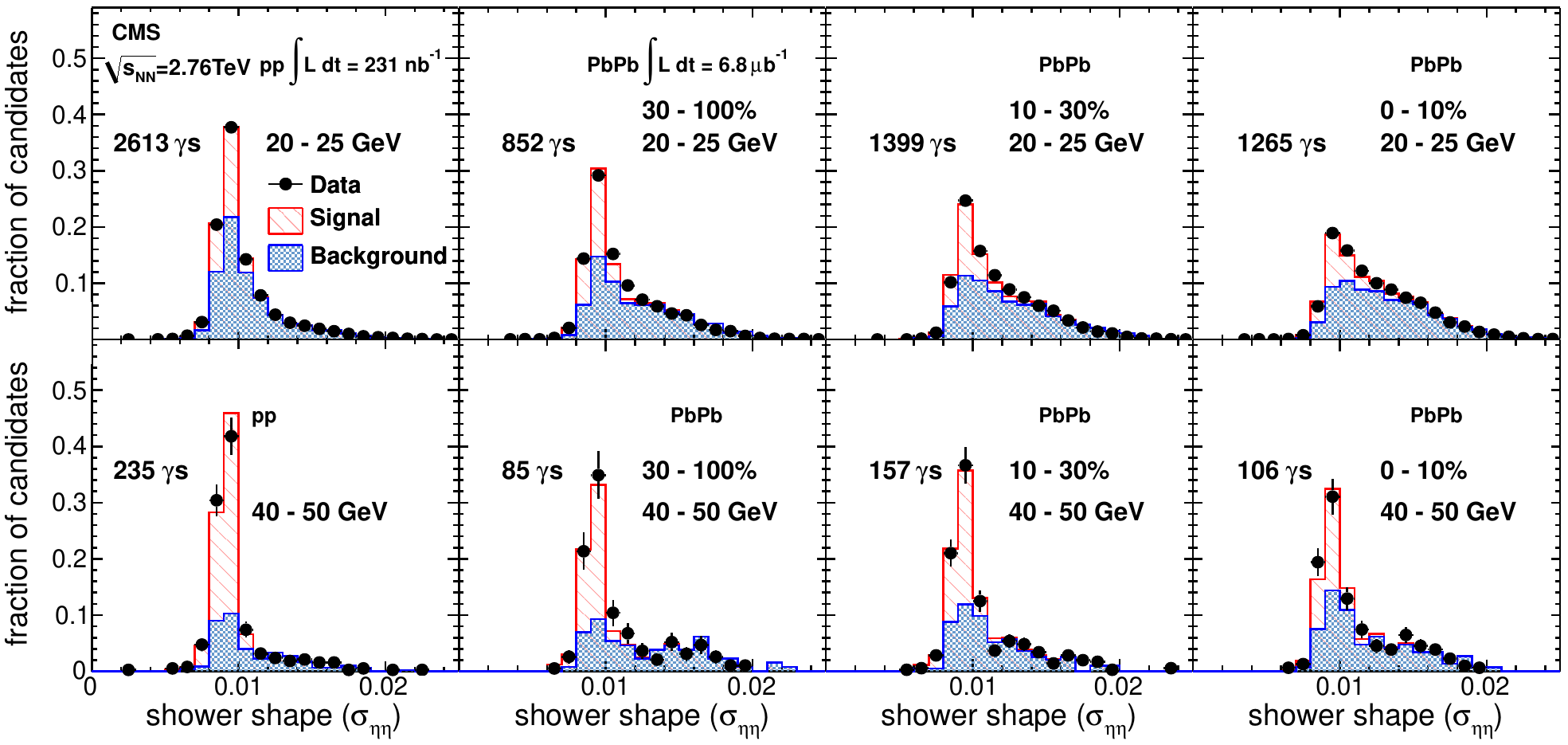}
    \caption{\label{fig:fit} Measured shower-shape $\sigma_{\eta \eta}$ distribution for photon candidates with
   $E_{\rm T}^\gamma = \mbox{20--25}$ GeV and 40--50GeV in pp (2 left plots) and PbPb collisions for 3 different centrality ranges. The extracted numbers of isolated photons are shown in the figure. The fit result (red line), signal (red-hatched histogram) and background components (blue shaded histogram) are also shown. }
  \end{center}
\end{figure*}

The selection criteria described above yield a relatively pure sample of isolated photons.
However, there are still non-prompt photons, such as those from isolated $\pi^0$s and $\eta$s that are carrying a large fraction of the parent fragmenting parton energy, which can pass the experimental isolation requirement. The remaining decay photon backgrounds are estimated using a two-component fit of the shape of the electromagnetic shower in the calorimeter and separated from the signal on a statistical basis. The topology of the energy deposits can be used as a powerful tool to distinguish the signal from the background by making use of the fine $\eta$ segmentation of the electromagnetic calorimeter.
The shower shape is characterized by a transverse shape variable $\sigma_{\eta \eta}$, defined as a modified second moment of the
electromagnetic energy cluster distribution around its mean $\eta$ position:
\begin{eqnarray}
\label{sieieFormula}
\sigma_{\eta \eta}^2 = \frac{\sum_i w_i(\eta_i-\bar{\eta})^2}{\sum_i w_i},~~~~ w_i = \mathrm{max}(0, 4.7 + \ln \frac{E_i}{E}),
\end{eqnarray}
where $E_i$ and $\eta_i$ are the energy and position of the $i^{th}$ crystal in a group of $5\times 5$ crystals
centered on the one with the highest energy, $E$ is the total energy of the crystals in the calculation and $\bar{\eta}$ is the average
$\eta$ weighted by $w_i$ in the same group~\cite{CMSppPhoton}. Clusters produced in hadron decays
tend to have larger $\sigma_{\eta  \eta}$ mean and a wider $\sigma_{\eta  \eta}$ distribution, while isolated photons tend to have a smaller mean value of $\sigma_{\eta \eta}$ and a narrow distribution.

The isolated prompt photon yield is estimated with a binned maximum likelihood fit to the transverse shower shape distribution with the expected signal ({\sc pythia+data}) and data-driven background components for each event centrality and $E^\gamma_{\rm T}$ interval~\cite{CMSpbpbPhoton}. The results of the two-component fit of the shower-shape distributions are shown in Fig.~\ref{fig:fit}. The remaining electron contribution estimated from data is subtracted from the fit result to extract the raw signal yields $(N^{\gamma}_{\rm raw})$. To correct the effect of the photon energy resolution, a bin-by-bin correction($U$) is also applied to the raw signal yields to obtain the final number of isolated photons. The $E_{\rm T}$-differential photon yield per event is defined as:
\begin{eqnarray}
 \frac{dN^{\gamma}_\mathrm{PbPb}}{dE^\gamma_{\rm T}} = \frac{N^{\gamma}_{\rm raw}}{U\times\epsilon \times f_{\rm cent}\times N_{\rm MB} \times \Delta E^\gamma_{\rm T}},
\end{eqnarray}
where $N_{\rm MB}$ is the number of sampled MB PbPb events, $f_{\rm cent}$ is the fraction of events in each centrality bin, and $\epsilon$ is the efficiency
of the photon identification. 

\begin{figure}[tb]
\begin{center}
\includegraphics[width=0.50\textwidth]{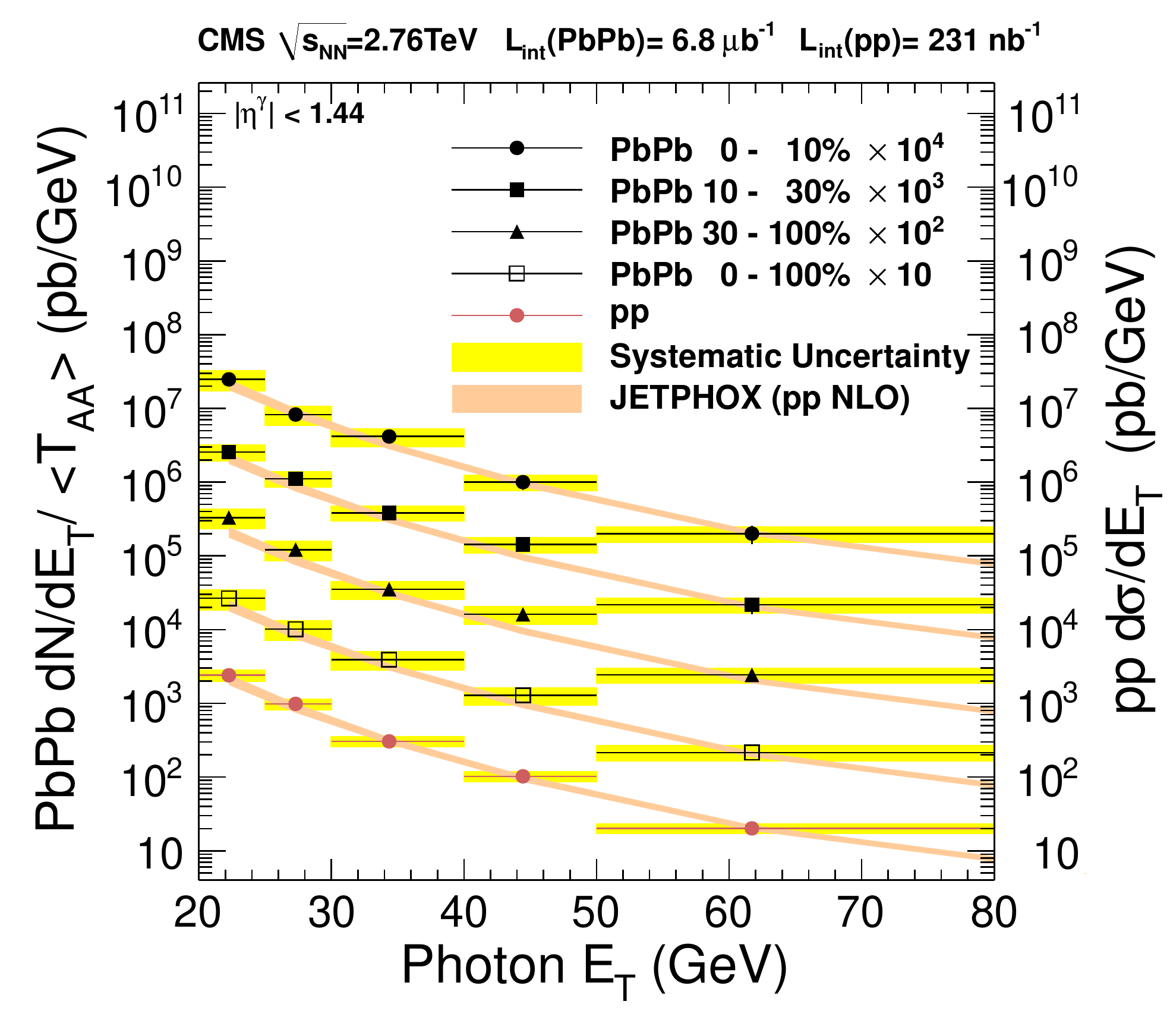}
\includegraphics[width=0.475\textwidth]{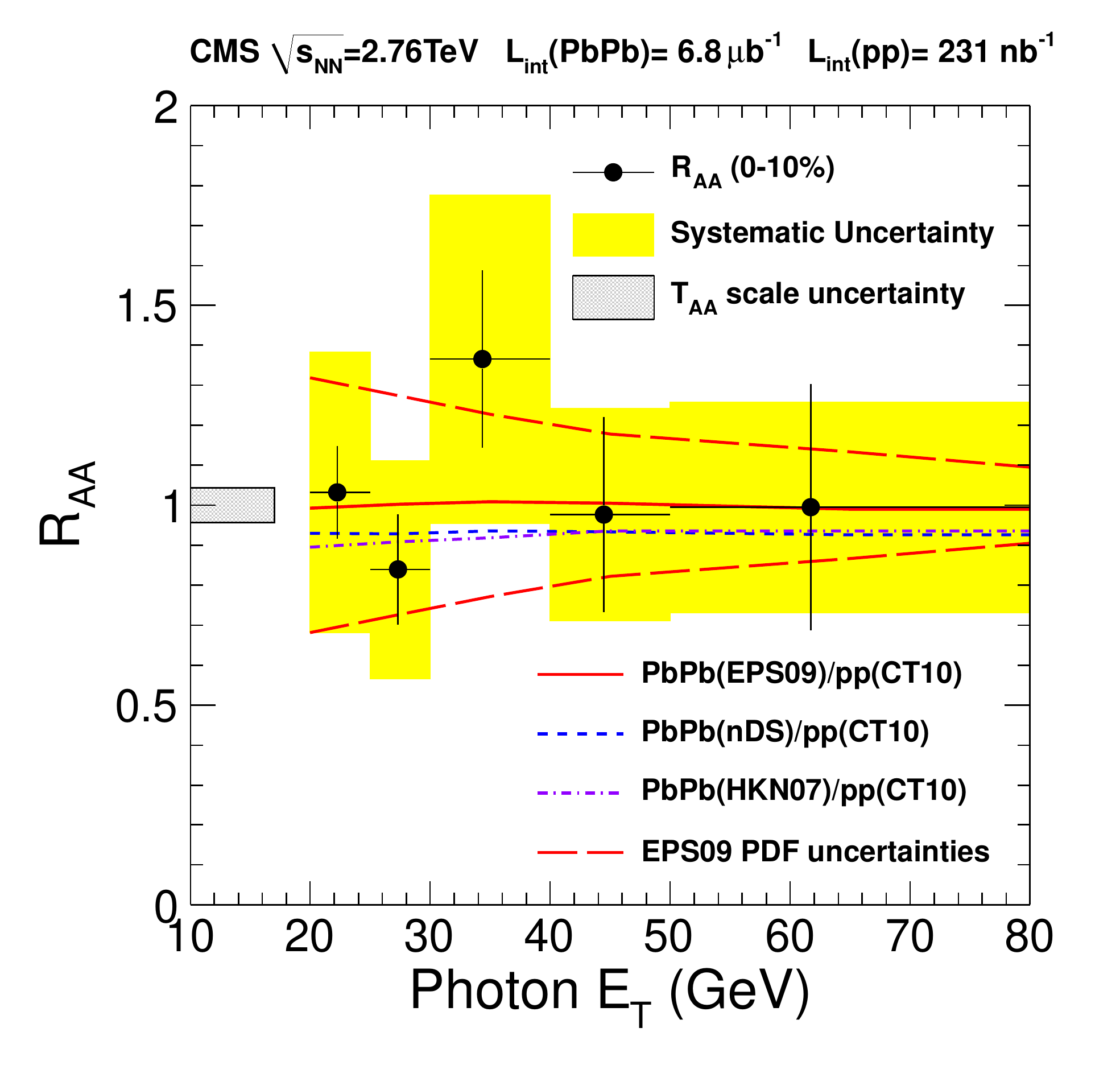}
\caption{\label{fig:dndet} (Left) Isolated photon spectra measured as a function of $E^\gamma_{\rm T}$ for 0--10\%, 10--30\%, 30--100\%,
  0--100\% PbPb collisions (scaled by $T_{\rm AA}$) and pp collisions at 2.76TeV. The horizontal bars indicate the bin width. The total systematic uncertainty is shown as a yellow box at each $E_{\rm T}$ bin. The results are compared to the NLO {\sc{jetphox}} calculation shown as a pink band. The vertical error bars indicate the statistical uncertainty and the horizontal bars reflect the bin width. (Right) Nuclear modification factor $R_{\rm AA}$ as a function of the photon $E_{\rm T}$ measured in the 0--10\% most central PbPb collisions at 2.76 TeV.}
\end{center}
\end{figure}

The total systematic uncertainties are 14--16\% for pp and 22--30\% for PbPb collisions which are dominated by the uncertainty on the background modeling. In order to compare the isolated photon cross sections in pp and PbPb collisions, a scaling factor, the nuclear overlap function $T_{\rm AA}$, is needed to provide proper normalization. This factor can be interpreted as the NN-equivalent integrated luminosity at any given PbPb centrality.  
Figure~\ref{fig:dndet} shows the pp cross sections and the PbPb $T_{\rm AA}$-scaled yields
compared to the {\sc{jetphox}} predictions obtained with the CT10 PDF. The pp and PbPb data are consistent with the NLO calculation at all photon transverse energies. The photon nuclear modification factor $R_{\rm AA}= dN^{\gamma}_{\rm PbPb}/dE^\gamma_{\rm T}/(T_{\rm AA}\times d\sigma^{\gamma}_{pp}/dE_{\rm T})$, is computed from the measured PbPb scaled yield and the pp differential cross section. Figure~\ref{fig:dndet} displays $R_{\rm AA}$ as a function of the isolated photon $E_{\rm T}$ for the 0--10\% most central PbPb collisions. The ratio is compatible with unity within the experimental uncertainties for all $E_{\rm T}$ values. 

In summary, the isolated photon spectra at $|\eta^\gamma|<1.44$ have been measured in pp and PbPb collisions at $\sqrt{s_{_{\rm NN}}}=2.76$ TeV with the CMS detector. The measured spectra are well reproduced by NLO pQCD calculations with recent PDFs for the proton and nucleus. It is consistent with the expectation that nuclear parton densities are not
significantly modified compared to the proton PDF in the explored kinematic range, dominated by high-$p_{\rm T}$ photons produced in parton-parton scatterings in the large-$Q^2$ and moderate parton fractional momentum $x$ region of the nuclear PDFs~\cite{Eskola:2009uj}. The measurement presented here establishes isolated photon production as a valuable perturbative probe of the initial state in heavy-ion collisions and provides a baseline for the study of in-medium parton energy loss in $\gamma$+jet events~\cite{CMSPhotonJet}.

\end{document}